\documentclass[manuscript]{aastex}

\shorttitle{Asymmetric CME-CME interaction}
\shortauthors{Temmer et al.}

\usepackage{natbib}
\usepackage{graphicx}
\usepackage{amsmath}
\usepackage{rotating}

\begin{document}
\title{Asymmetry in the CME-CME interaction process for the events from 2011 February 14--15}

\author{M. Temmer\altaffilmark{1}, A.~M. Veronig\altaffilmark{1}, V. Peinhart\altaffilmark{1}}
\affil{$^1$Kanzelh{\"o}he Observatory-IGAM, Institute of Physics, University of Graz, Universit{\"a}tsplatz 5, 8010 Graz, Austria}
\and
\author{B. Vr{\v s}nak\altaffilmark{2}}
\affil{$^2$Hvar Observatory, Faculty of Geodesy, University of Zagreb, Ka\v{c}i\'{c}eva 26, HR-10000 Zagreb, Croatia}

\date{Received/Accepted}

\begin{abstract}
We present a detailed study of the interaction process of two coronal mass ejections (CMEs) successively launched on 2011 February 14 (CME1) and 2011 February 15 (CME2). Reconstructing the 3D shape and evolution of the flux ropes we verify that the two CMEs interact. The frontal structure of both CMEs measured along different position angles (PA) over the entire latitudinal extent, reveals differences in the kinematics for the interacting flanks and the apexes. The interaction process is strongly PA-dependent in terms of timing as well as kinematical evolution. The central interaction occurs along PA-100$^{\circ}$, which shows the strongest changes in kinematics. During interaction, CME1 accelerates from $\sim$400~km~s$^{-1}$ to $\sim$700~km~s$^{-1}$ and CME2 decelerates from $\sim$1300~km~s$^{-1}$ to $\sim$600~km~s$^{-1}$. Our results indicate that a simplified scenario like inelastic collision may not be sufficient to describe the CME-CME interaction. Magnetic field structures of the intertwining flux ropes as well as momentum transfer due to shocks play an important role in the interaction process. 
 \end{abstract}


\maketitle

\section{Introduction}

The kinematical behavior of isolated coronal mass ejections (CMEs) can strongly differ compared to a series of CMEs. Successive eruptions from the Sun cause a change of the interplanetary environment (e.g. density, speed) and complex magnetic structures may form during CME-CME interaction. Observable features of CME-CME interaction are restricted to white-light and radio data. The CME kinematics is found to clearly change during the collision of two CMEs, resulting in either strong deceleration \citep[e.g.,][]{temmer12} or acceleration \citep[][super-elastic collision]{shen12}. Strong deceleration in the early kinematical evolution was also reported for CMEs that run into strong overlying magnetic fields \citep{temmer10}. It is generally accepted that the shape and orientation of the magnetic flux rope embedded in a CME plays an important role for the further propagation and evolution in interplanetary space \citep[e.g.][]{schmidt04, lugaz05, lugaz13}.

The kinematical behavior for a chain of CME events from 2011 February 14 to 2011 February 15 and their production of a Forbush decrease is described by \cite{maricic13}. They find evidence for a gradual momentum transfer between the slower and the faster CME, resulting in the deceleration of the faster and the acceleration of the slower one \citep[see also][]{farrugia04}. The momentum transfer may take place through a CME-driven shock. This interpretation is supported by numerical simulations showing that fast shocks may travel through slower CME fronts ahead \citep[see e.g.,][]{lugaz09,liu12}.

The current analysis is an extension of the study by \cite{maricic13}. We investigate in detail the interaction process of the two consecutive CMEs launched in a time window of about 8 hours during 2011 February 14--15. Using spherically de-projected Heliospheric Imager data in combination with a 3D reconstruction of the magnetic flux ropes, we track the frontal structures of both CMEs over their entire latitudinal extent in order to study asymmetries in the interaction process. We also investigate dynamic radio spectra in the decametric/hectometric domain to study the nature of the interaction process.

\section{Data and Methods}

\begin{figure}
	\epsscale{0.8}
		\plotone{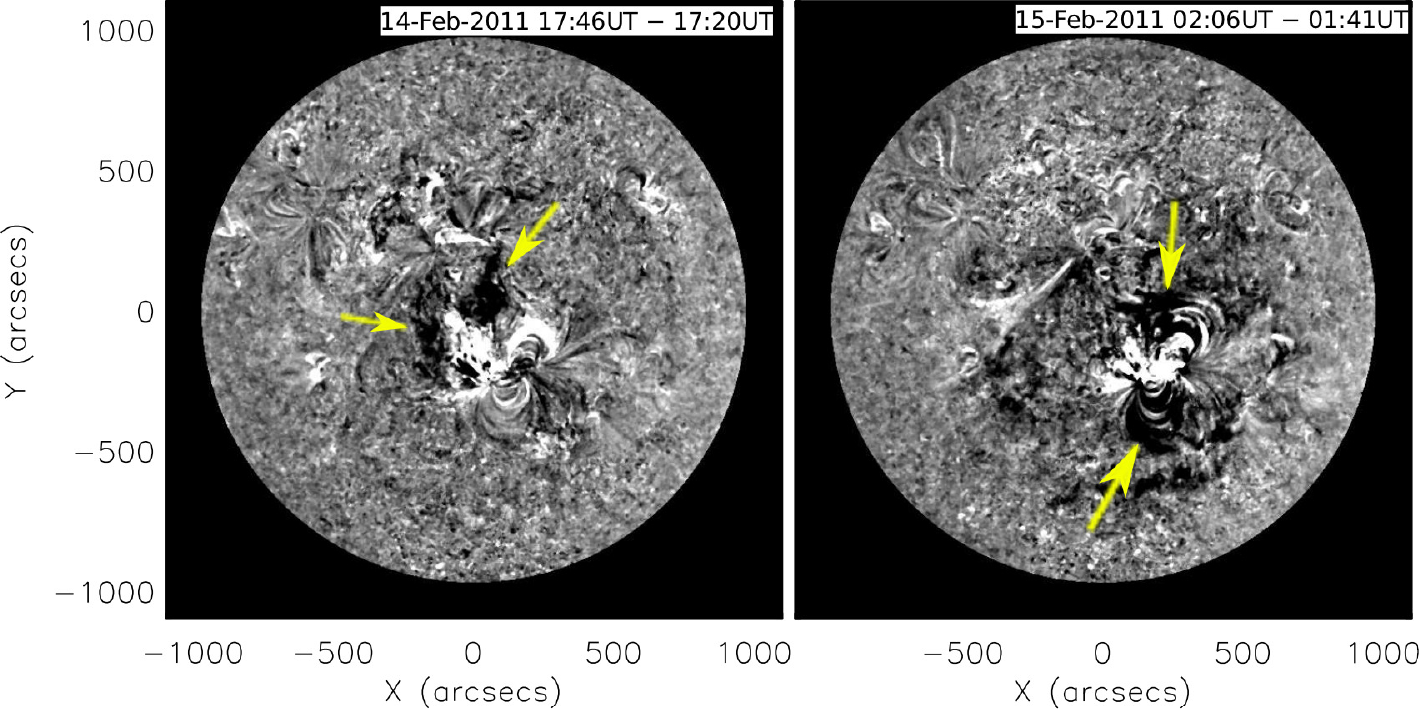}
	\caption{Proba2/SWAP 174{\AA} base-difference images showing weak dimming regions close to the active region (marked with yellow arrows) associated with the two CMEs under study. }
	\label{event}
\end{figure}

Figure~\ref{event} shows base difference images from Proba2/SWAP in the EUV 174\AA~passband \citep{berghmans06}. Around active region AR~11158 weak dimming regions are observed, evidencing the source region from where the CME events under study are launched. The slower CME (hereafter CME1) is associated to an M2.2 flare on 2011 February 14 17:20~UT (location: S20W04) and the faster CME (hereafter CME2) to an X2.2 flare on 2011 February 15 at 01:44~UT (location: S20W10).

To clearly identify the successively launched CMEs as colliding events and to study the interacting flux ropes in detail, we apply the 3D reconstruction technique from the forward model by \cite{thernisien06}. Using three different vantage points from which the CMEs are observed, we are able to determine the width and direction of motion of the CMEs. The forward model is based on the graduated cylindrical shell model, assuming that the idealized flux rope resembles the body of a CME. We note that the CME density envelope, as observed in white-light data, most probably covers the flux rope and the shock-sheath region \citep[see e.g.,][]{vourlidas13} and that the 3D reconstruction, by using a simplified geometry of the CME, cannot fully cover the complex morphology of the CME. For our purposes we define the best fit of the GCS model to the observations when the GCS flux rope boundary matches the outer edge of the CME structure and avoids compressed streamer structures. For this we fit the model to contemporaneous image triplets from the Solar Terrestrial Relations Observatory \citep[COR2 from STEREO-A and STEREO-B; see][]{howard08} and the Solar and Heliospheric Observatory \citep[SoHO/LASCO; see][]{brueckner95}. From the reconstruction we derive the source location of the CME on the Sun (back-projecting the CME apex along a straight line normal onto the solar surface), the direction of motion, the edge-on and face-on width of the CME body, and the propagation speed within the COR2 field-of-view (FoV), ranging from 2.5 to 15~R$_{\odot}$.

Using STEREO-A's Heliospheric Imager data HI1-A we measure for each CME the elongation of the outermost bright structure, i.e., the CME front, along different position angles (PAs) in running difference images with a cadence of 40 minutes. The off-pointed and wide-field HI1-A images are transformed to a Sun-centered polar coordinate system mapping the elongation angle along the radial direction and the PA around the azimuth \cite[see e.g.,][]{sheeley08,sheeley10}. For precise measurements of the two CME frontal structures we use direct images as well as time-elongation maps \citep[so called JMaps;][]{sheeley99} separately constructed for different PAs. The measured elongation angle is converted into radial distance using the harmonic mean method by assuming that we observe the tangent to the CME apex which is idealized by a circular front \citep[see][]{lugaz05,lugaz09}. For the conversion we take the direction of motion of each CME derived from forward fitting. For the kinematical study we apply a simple technique by calculating the linear regression curve to the derived distance-time data and its residuals (measured minus fitted values). Systematic residuals indicate phases of speed increase or decrease with respect to the mean constant speed.

The CME mass is derived from the brightness increase caused by the CME in COR2-A base-difference images. The intensity values of the pixels are converted into mass using the Thomson scattering formulation assuming that the CME has a composition of completely ionized hydrogen and 10\% He and that all electrons are located on the spacecraft plane-of-sky \citep{colaninno09}. This assumption is reasonable due to the position of the source region of the CMEs close to central meridian and the large separation angle of STEREO-A with Earth ($\sim$87$^{\circ}$).

For a more complete picture of the interaction process we study in addition to white-light data, dynamic radio spectra with 1-minute resolution from the Wind/WAVES experiment \citep{bougeret98} together with the interplanetary density model from \cite{vrsnak04}.

\section{Results}

Figure~\ref{event} shows close to AR~11158 weak dimming regions (marked with yellow arrows) which are associated with each of the two CME events, providing us with a first imprint of the CME's size and direction of motion. The dimmings reveal that both CMEs are initiated close to the solar disk center. CME1 is located more to the North of the AR and CME2 is directed more to the South, being centered at AR~11158. The dimmings for the second CME are more significant, both by amount and extent.

\begin{figure}
	\epsscale{0.7}
		\plotone{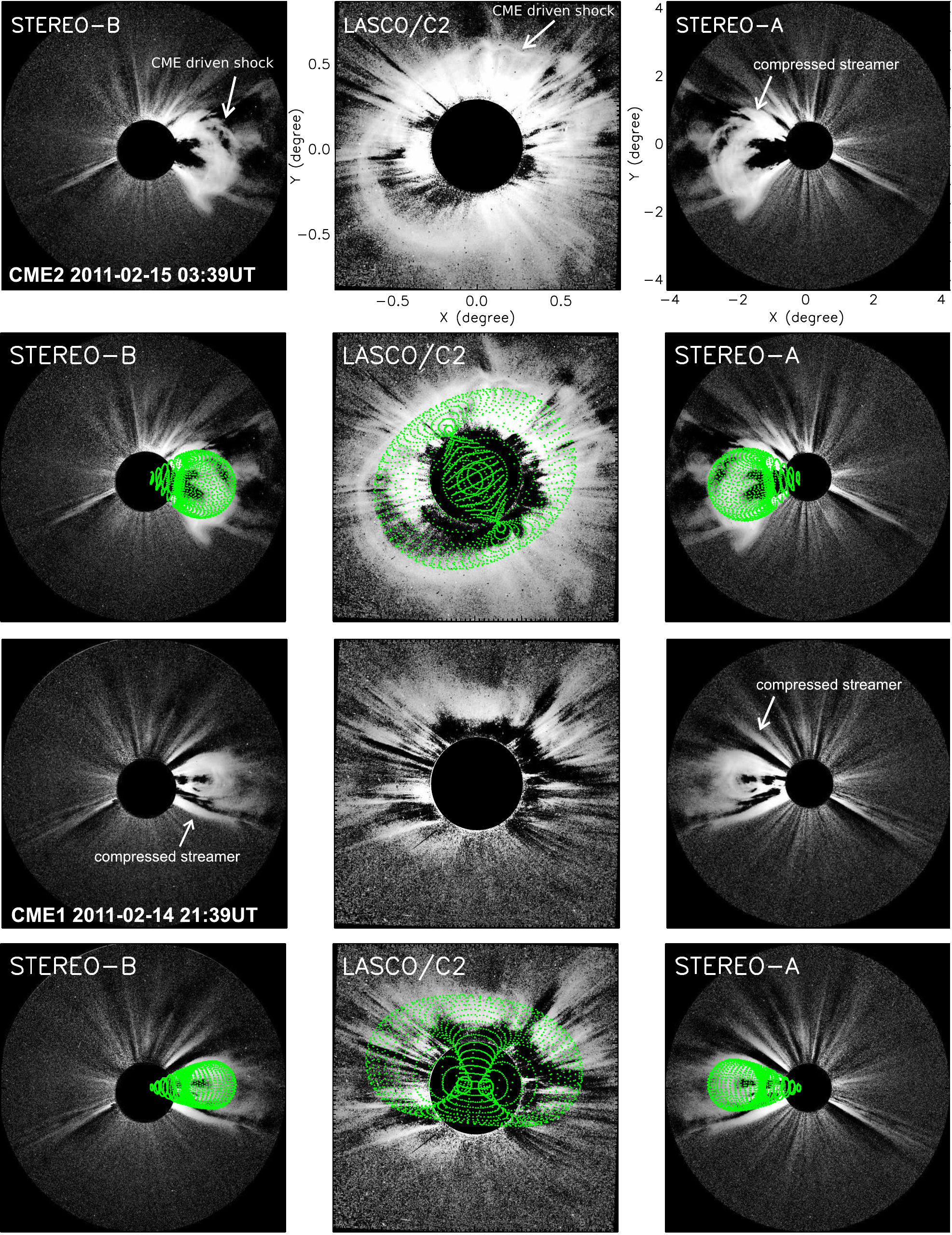}
	\caption{Contemporaneous image triplets from STEREO-A (right), STEREO-B (left), and LASCO (middle) data showing CME1 (bottom panels) and CME2 (top panels). Results from the graduated cylindrical shell model are shown with the green mesh for each CME.}
	\label{CME1a2}
\end{figure}

Figure~\ref{CME1a2} presents the image triplets (STEREO-A, LASCO, STEREO-B) on which the flux rope resulting from the forward model is applied (top panels show CME2 and bottom panels CME1). For comparison, the middle-row panels show model results for both CME bodies as obtained at the distance of 10.5~R$_{\odot}$. We derive that the center of the apex of CME1 back-projected onto the solar surface is located at W00S02, for CME2 it is found at W00S07. These results have an error of $\pm$5$^{\circ}$ and are due to uncertainties in the determination of the boundary shell of the CME, since it is sometimes hard to distinguish between flux rope body and compressed plasma (cf.\ top right panel in Figure~\ref{CME1a2}). We derived an average speed over the COR2 FoV for CME1 of $\sim$390~km/s and $\sim$1020~km/s for CME2, with maximum speeds of 450~km/s and 1300 km/s, respectively, indicating that CME2 strongly decelerates early in its evolution \citep[see also][]{maricic13}. The axial orientation of CME1 has a tilt of $-$8.0$^{\circ}$, i.e.\ rotated about 8$^{\circ}$ clockwise out of the ecliptic plane, and for CME2 $+$26.2$^{\circ}$. Calculating the volume of the idealized flux rope \citep{thernisien06}, we obtain 1.4$\cdot$10$^{35}$~cm$^{3}$ for CME1 and 2.1$\cdot$10$^{35}$~cm$^{3}$ for CME2. The derived mass is typical for CMEs \citep[see e.g.,][]{vourlidas10,bein13} and similar for both events (CME1: $4.7\cdot10^{15}$~g and CME2: $6.4\cdot10^{15}$~g). Based on the volume and the mass we also estimate the density, which is 3.5$\cdot$10$^{-20}$~g~cm$^{-3}$ for CME1 and 3.1$\cdot$10$^{-20}$~g~cm$^{-3}$ for CME2. The obtained 3D parameters of each CME are also given in Table~\ref{table1}. In summary, visual inspection as well as the derived 3D model parameters clearly outline that both CMEs propagate more or less in the same direction and are of similar density, have somewhat different axial orientations and shapes, and significantly different speeds.

\begin{figure}
	\epsscale{1.}
		\plotone{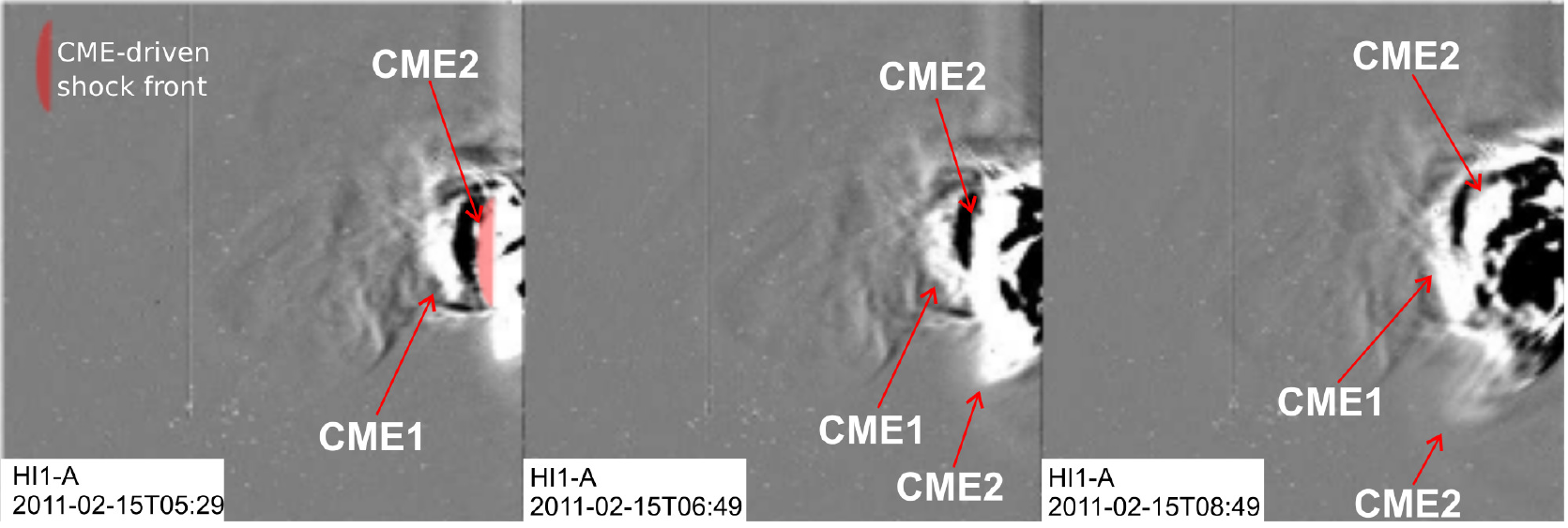}
	\caption{HI1-A running difference images showing the interaction of CME1 and CME2. Arrows mark the fronts of CME1 and CME2. CME2 forms a bulk in the southern region where no interaction takes place. }
	\label{fig3}
\end{figure}

Figure~\ref{fig3} shows details of the CME-CME interaction as it occurs in the field-of-view of HI1-A. At 05:29~UT both CME fronts can be well distinguished. At 06:49~UT the frontal structure of CME2 changes, providing evidence for the ongoing interaction process. Later on, frontal parts of CME1 may be hidden by the increase in intensity as well as due to line-of-sight effects as CME2 encounters CME1. The environmental conditions (speed, magnetic field, density) in which CME2 propagates obviously change over its latitudinal extent. For the southern part of CME2 a side-lobe forms which quickly weakens in intensity compared to northern latitudes. We note that the southern part of the CME (PA-125) might have been swept radially forward by the fast solar wind coming from the coronal hole in the south \citep[see e.g.,][]{manchester04}. 
 
\begin{figure}
	\epsscale{0.7}
		\plotone{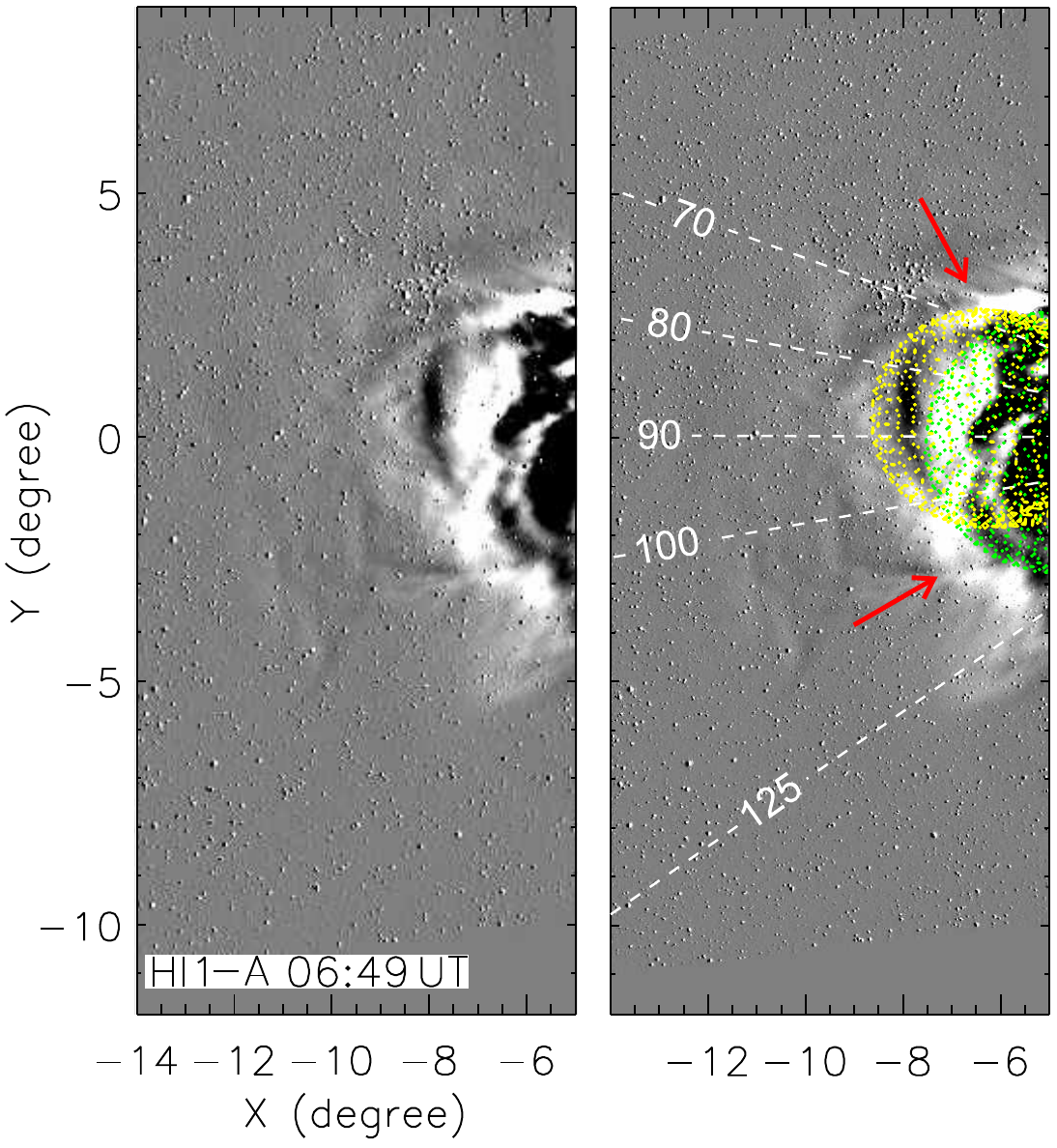}
	\caption{Spherically de-projected HI1-A difference image at 06:49~UT (left) together with the results from the 3D forward model for the respective time (right). The GCS flux rope of CME1 is shown as yellow mesh and of CME2 as green mesh. Red arrows mark the location where the boundaries of both flux ropes are assumed to interact. The accompanying movie shows the entire series of spherically deprojected images covering the time range 2011 February 15 01:29~UT until 2011 February 16 05:29~UT. For better guidance, red arrows mark the front of CME2. }
	\label{inter}
\end{figure}

Assuming self similar expansion, we extrapolate the derived flux rope bodies for both CMEs (cf.\ Figure~\ref{CME1a2}) to the field-of-view of HI1-A. In Figure~\ref{inter}, the result is shown at the time 06:49~UT. We derive that the flux ropes of CME1 and CME2 are basically located over PA-70--100$^{\circ}$, hence, the southern part of CME2 along PA-125$^{\circ}$ is not influenced in its propagation by CME1. We can also identify increased intensity, indicating compression, for the regions where the boundaries of both GCS flux ropes are assumed to interact (marked by red arrows in Figure~\ref{inter}).

The left panels of Figure~\ref{resid} show the distance-time data for both CMEs derived from elongation measurements (HI1-A FoV) along different PAs that are converted into radial distance by the harmonic mean method. Error bars (not shown) lie in the range of about $\pm$50~km~s$^{-1}$ and are due to uncertainties in the identification of the outermost bright CME structure, caused by the intensity drop off. We calculate for each distance-time profile a linear fit to obtain the average interplanetary propagation speed (exemplary outlined for PA-100$^{\circ}$). The average speeds along PA-70--100$^{\circ}$ are $v_{\rm CME1}$=440$\pm$40~km~s$^{-1}$ and $v_{\rm CME2}$=530$\pm$50~km~s$^{-1}$. Comparing these results with those obtained from the forward model for the entire flux rope body within the COR2 FoV, we obtain that the mean speed of CME1 increased while CME2 strongly decelerated. For the kinematical evolution of CME2 along PA-125$^{\circ}$ we obtain an average speed of $\sim$960~km~s$^{-1}$ which is comparable to the result derived from the forward model. The encounter between CME1 and CME2 can be inferred by the approaching kinematical curves at 2011 February 15 for PA-80$^{\circ}$ and PA-100$^{\circ}$ around 08~UT, along PA-90$^{\circ}$ around 12~UT, and for PA-70$^{\circ}$ after 16~UT indicating different times of encounter over the latitudinal extent of the frontal structures of CME1 and CME2.

The right panels of Figure~\ref{resid} show the residuals for the CME kinematics derived, i.e.\ the difference between the distance-time measurements and the fitted values. Residuals distributed randomly around zero indicate a more or less constant speed of the CME propagation. Systematic residuals over certain periods indicate deviations from a constant speed, either increasing or decreasing. The most obvious changes in the residuals (of both CMEs) are obtained over the directions PA-80--100$^{\circ}$. We derive for CME1 a clear drift from positive to negative values, i.e.\ deceleration, over the time range 2011 February 14 22:00~UT -- 2011 February 15 08:45~UT, a reverse trend, i.e.\ acceleration, occurs during the period 08:45--14:00~UT followed again by a deceleration. Most distinctly the changes are observed along PA-100$^{\circ}$, giving evidence for the central interaction along this direction. Quantifying these systematic changes measured along PA-100$^{\circ}$, we derive average speeds of about 400~km~s$^{-1}$, 700~km~s$^{-1}$, and 500~km~s$^{-1}$ for the three periods. In comparison, along PA-70$^{\circ}$ no strong deviations from the average speed are observed as the CMEs evolve within the FoV of HI1-A. For PA-80$^{\circ}$ we also show the trailing edge of CME2 which clearly shows less variation in its speed compared to the frontal part of CME2. Inspecting the distance-time data and residuals for different PAs shows that the interaction process is strongly PA-dependent in terms of timing as well as kinematical evolution.

Figure~\ref{Bflare} shows the dynamic radio spectra from WIND/Waves receivers RAD1/RAD2 covering the frequency range 14~MHz--200~kHz, approximately corresponding to a distance range from 2.5~R$_{\odot}$ to 30~R$_{\odot}$. During 2011 February 15 02:00--07:00~UT, i.e.\ shortly after the launch of CME2, we observe the fundamental and harmonic plasma emission of a type II burst due to a shock that accelerates electrons in its upstream region. To determine its propagation speed we fit the visible part of the type II burst using the interplanetary density model by \cite{vrsnak04}, and derive a constant speed of about $v$=1100~km~s$^{-1}$ which matches well with the speed of the leading edge of CME2 in the FoV of COR2 (cf. Table~\ref{table1}). Several enhancements of the type II radio burst are observed (marked by white arrows in Figure~\ref{Bflare}). The first one occurs at around 02:40~UT in between the fundamental and harmonic band of the type II burst, a second one intersects the harmonic band at around 03:20 followed by an intense continuum-like radio emission covering the frequency range 400~kHz--1.0~MHz (corresponding to a distance range of $\sim$10--18~R$_{\odot}$) during 04:45--05:30~UT. After the launch of the second CME, frequent type III bursts are observed related to flare emission from AR~11158. 
 
\begin{table*}
	\centering
	\small
		\begin{tabular}{|c|c|r|c|r|c|c|c|c|c|}
		\hline  event & position  & tilt          & time  & speed & width  & width & volume & mass & density \\
                   no.\       &  &   [$^{\circ}$]   &    [UT]    &[km/s] & fo [$^{\circ}$] & eo [$^{\circ}$] & [cm$^{3}$] & [g] & [g~cm$^{-3}$] \\
\hline \hline
 			CME1 & W00S02 & $-$8.0 & 21:39 2011-02-14 &  390 & 100 & 25 & 1.4$\cdot$10$^{35}$ & 4.7$\cdot$10$^{15}$ & 3.5$\cdot$10$^{-20}$ \\
            CME2 & W00S07 & 26.2 & 03:39 2011-02-15 & 1020 & 90 & 60 & 2.1$\cdot$10$^{35}$ & 6.4$\cdot$10$^{15}$ & 3.1$\cdot$10$^{-20}$ \\
		 \hline
	
		\end{tabular}
		\caption{3D CME parameters derived from forward modeling to the STEREO COR2-A, COR2-B and SOHO LASCO C3 observations at a CME distance of 10.5~R$_{\odot}$. The volume is calculated from the derived face-on (fo) and edge-on (eo) widths, assuming an idealized flux rope with the shape of ``two ice cones plus half donut'' \citep[see also][]{thernisien06}. The speed is an average over the FoV of COR2.}
		\label{table1}
\end{table*}

\section{Discussion and Conclusion}

We followed in detail the CME-CME interaction event from 2011 February 14--15. Applying a graduated cylindrical shell model \citep{thernisien06} to the multi-spacecraft coronagraphic data we derive that both CMEs head into the same direction, with different speed, size and axial orientation. The front of CME1 is less bright than CME2 and its structure is intermittent and increased in intensity close to PA-70$^{\circ}$ and PA-100$^{\circ}$, which matches with the location where both flux ropes are supposed to interact.

\begin{figure}
	\epsscale{0.8}
		\plotone{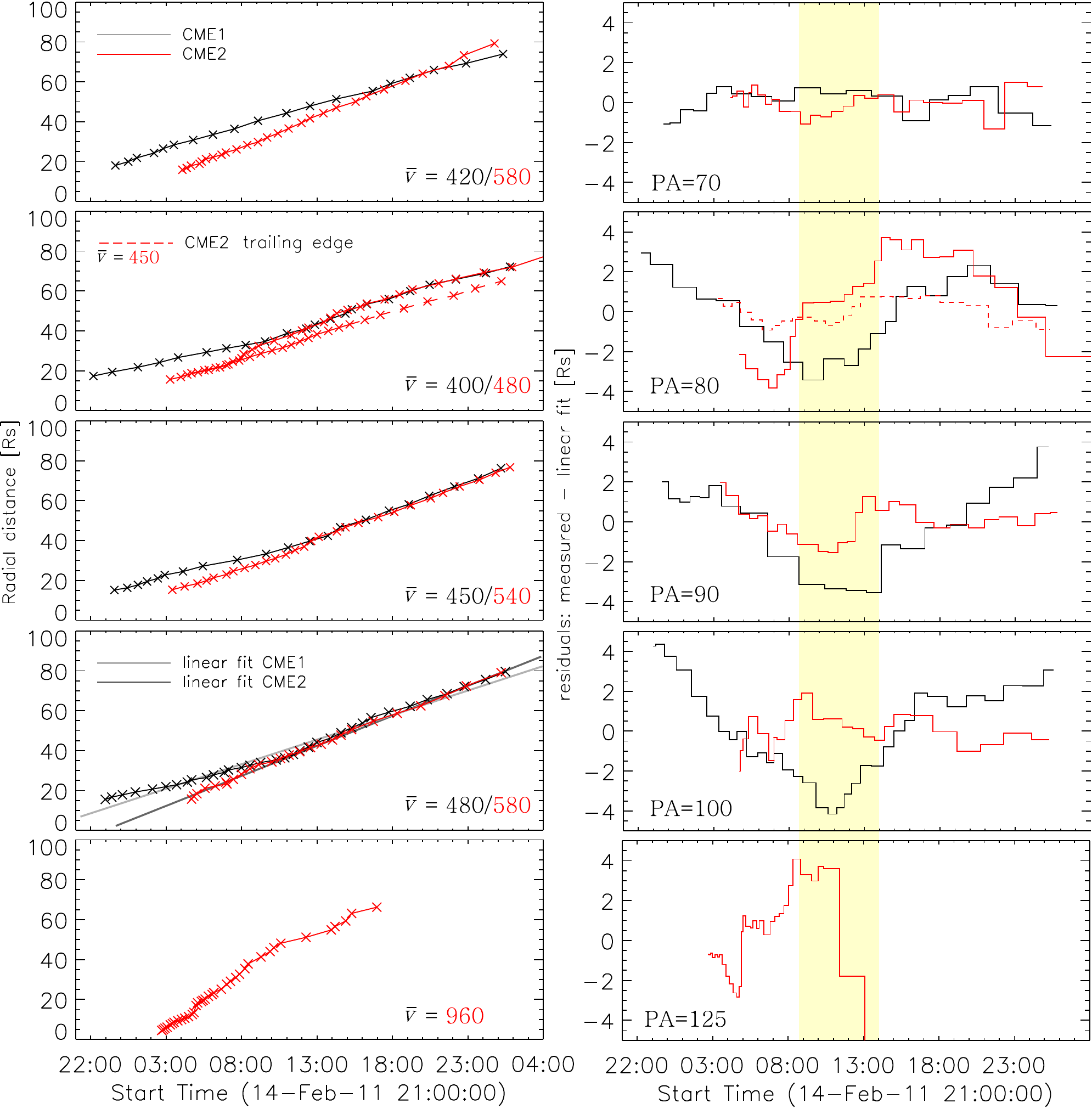}
	\caption{De-projected radial distances (left panels) and residuals showing measured minus fitted values (right panels) over time. Measurements are taken along different PAs [70, 80, 90, 100, and 125$^{\circ}$] from spherically de-projected HI1-A data for the fronts of CME1 (black lines) and CME2 (red lines; in addition for PA-80$^{\circ}$ the trailing edge is given as dashed red line). For PA-100$^{\circ}$ we show the linear fit to the distance-time data for both CMEs. This should help properly interpreting the derived residuals. Shaded area in the right panels marks the time of CME encounter for PAs 80--100$^{\circ}$. }
	\label{resid}
\end{figure}

\begin{figure}
	\epsscale{0.8}
	\plotone{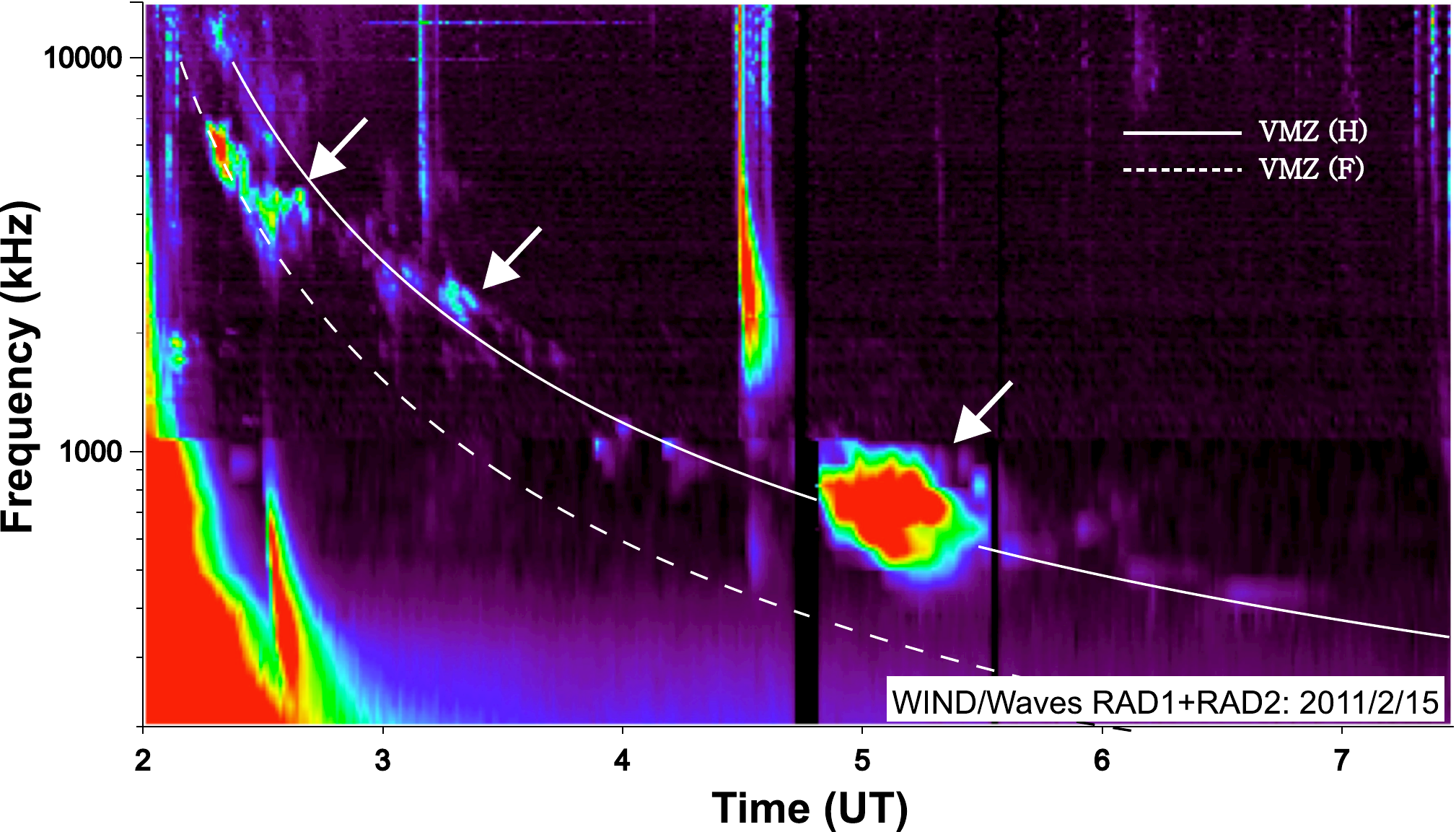}
		\caption{Dynamic radio spectra from WIND/Waves receivers RAD1/RAD2. The type II emission is fitted by the interplanetary density model by \cite{vrsnak04} drawn by white solid/dashed lines for the harmonic/fundamental plasma emission. The white arrows indicate enhancements of the type II burst. }
	\label{Bflare}
\end{figure}

From the kinematical curves measured along different PAs (cf.\ Figure~\ref{resid}), we derive different speeds and different times for the encounter between the fronts of both CMEs. This might be a consequence of the different axial orientations of the flux ropes estimated to deviate from each other by $\sim$35$^{\circ}$ and would indicate that the interacting magnetic field structures are decisive for the evolution of the interaction process. We would like to note that the strong kinematical changes measured for the frontal structures along PA-80--90$^{\circ}$, affect both CMEs at about the same time. Since CME2 propagates in the wake of CME1, changes in the speed of CME1 most likely affect the kinematics of CME2. Inspecting the behavior of the southern part of CME2 along PA-125$^{\circ}$, we obtain that its frontal structure is highly deformed and propagates with much higher speed (960~km~s$^{-1}$) than the frontal structure measured along PA-70--100$^{\circ}$ (530$\pm$50~km~s$^{-1}$). These results give further evidence on the latitudinal range of influence of the flux rope of CME1, associated with strong changes of the ambient density, speed and magnetic flux. Deformation of CME fronts (concave-outward shape, pancaking) due to different solar wind speeds in low and high latitudinal regions is reported by e.g., \cite{russell02}, \cite{riley04}, and \cite{liu06}.

For the direction of central collision along PA-100$^{\circ}$, we perform the exercise of assuming the simplified scenario of an inelastic collision between two bodies of similar mass (cf.\ Table~\ref{table1}). To push CME1 from 400 to 700~km~s$^{-1}$, as observed around $\sim$9~UT, CME2 would need a speed of at least 1000~km~s$^{-1}$. CME2 reached a maximum speed of $\sim$1300~km~s$^{-1}$ within the FoV of COR2, but decelerated during the assumed time of collision already to $\sim$600~km~s$^{-1}$. This indicates that CME2 may influence CME1 much earlier than suggested by the white-light data. Similar results are found for the CME-CME interaction from 2010 August 1, for which a strong deceleration of the faster CME was observed a few hours before its leading edge showed in the heliospheric images the merging with the slower CME ahead \citep{temmer11}. \cite{maricic13} concluded that the CMEs are of finite thickness, thus to a finite ``signal'' travel time the interaction process may start well before changes are observed at the leading edge of CME1, most likely through transfer of momentum \citep[see e.g.,][]{farrugia04}. A recent paper by \cite{lugaz13} identifies a number of problems in determining the type of collision between two CMEs among others due to changes in the mass of the CMEs during evolution \citep[see also][]{bein13} as well as ongoing perturbation hours after the collision.

Several enhancements of the radio type II burst associated to CME2 are observed and may give hint that the interaction process started as early as 2011 February 15 02:40~UT \citep[see e.g.,][]{gopalswamy01}. However, this first enhancement appears about 1 hour after the launch of CME2, much too early for CME2 or the shock of CME2 ($v_{\rm shock}\approx$ 1100~km~s$^{-1}$) to reach CME1 which is roughly at a distance of 20~R$_{\odot}$ at that time. Recent findings show that radio enhancements in type II bursts are frequently related to shock-streamer interactions \citep{shen13}. As a large part of CME2 (PA-70--100$^{\circ}$) propagates in the rear of CME1, the radio enhancements could be a consequence of the interaction between the shock of CME2 and the streamer-like post-eruption current sheet formed behind CME1. Interestingly, the enhancements are found to be associated to type III bursts which are all related to flares emitted from AR~11158, the same AR where both CMEs are launched from. Type III radio bursts subsequently observed on 2011 February 15 (not shown), stemming from the same AR, stop at frequencies related to the downstream region of the extrapolated type II burst, as if it would be a barrier for particles entering the magnetic structure \citep[see also][]{macdowall89}. In a follow-up study we aim to make a more thorough analysis on the connection between the cutoff frequencies of type III and type II bursts as well as their relation to the process of CME-CME interaction. In conclusion, our observational data reveal only the consequences of CME-CME interaction that may not be sufficient to fully describe the process of interaction. The shape and orientation of the magnetic structures of CME-CME events certainly play a key role.

\acknowledgements
We would like to thank K.-L. Klein for his valuable comments and fruitful discussions. M.T. greatly acknowledges the Fonds zur F\"orderung wissenschaftlicher Forschung (FWF): V195-N16. The research leading to these results has received funding from the European Commission's Seventh Framework Programme (FP7/2007-2013) under the grant agreement n$^{\circ}$~284461 [eHEROES].

\clearpage

\clearpage

\clearpage

\clearpage

\clearpage

\bibliographystyle{apj}

\end{document}